\documentclass[prd,nofootinbib,twocolumn,showpacs]{revtex4}
\usepackage{graphics}
\usepackage{bm}

\begin{document}

\title {The Effects of Multiplicative Noise in Relativistic Phase Transitions}

\author{Nuno\ D.\ Antunes}  \affiliation{
Center for Theoretical Physics, University of Sussex, \\
Falmer, Brighton BN1 9WJ, U.K.}
\author{Pedro\ Gandra} \author{Ray\ J.\ Rivers}
\affiliation{ Blackett Laboratory, Imperial College\\
London SW7 2BZ, U.K.}

\begin{abstract}
Effective stochastic equations for the continuous transitions of
relativistic quantum fields inevitably contain multiplicative
noise. We examine the effect of such noise in a numerical
simulation of a temperature quench in a 1+1 dimensional scalar
theory. We look at out-of-equilibrium defect formation and compare
our results with those of stochastic equations with purely
additive noise.
\end{abstract}


\

\maketitle

\section{Introduction}

Since phase transitions take place in a finite time, causality
guarantees that correlation lengths remain finite, even for
continuous transitions. Because of the universal presence of
causality, Kibble \cite{kibble1} and Zurek \cite {zurek1,zurek2}
suggested that it alone is sufficient to bound the size of
correlated domains after the implementation of a continuous
transition. There are several ways\cite{zurek2} of formulating
causality bounds, but they all depend on the fact that, as the
transition begins to be implemented, there is a maximum speed at
which the system can become ordered. For relativistic quantum
field theory (QFT) this is the speed of light whereas, for
superfluids, for example, it is the speed of second sound.

The argument is very general. Consider a system with critical
temperature $T_{c}$, cooled through that
temperature so that, if $T(t)$ is the temperature at time $t$, then $%
T(0)=T_{c}$. ${\dot{T}}(0)=T_{c}/\tau _{Q}$ defines the quench
time $\tau _{Q}$. Suppose that the adiabatic correlation length
$\xi _{ad}(t)=\xi _{ad}(T(t))$  diverges near $t=0$ as
\[
\xi _{ad}(t)=\xi _{0}\bigg|\frac{t}{\tau _{Q}}\bigg|^{-\nu }.
\]
\noindent The fundamental length scale of the system $\xi _{0}$ is
determined from the microscopic dynamics. Although $\xi _{ad}(t)$
diverges at $t=0$ this is not the case for the true
non-equilibrium correlation length $\xi (t)$, which can only
change so much in a finite time. Kibble and Zurek made the
assumption that the correlation length ${\bar{\xi}}$ of the fields
that characterizes the
onset of order is the equilibrium correlation length ${\bar{\xi}}=\xi _{ad}({%
\bar{t}})$ at some appropriate time ${\bar{t}}$.

For simple systems all estimates of ${\bar t}$ (the 'causal time')
agree, up to numerical factors approximately unity \cite{zurek2}.
Most simply, $\xi (t)$ cannot grow faster than $c(t)= c(T(t))$,
where $c(T)$ is the causal velocity at temperature $T$.[For
relativistic theories $c(T) = c$, constant, whereas for condensed
matter systems we typically have critical slowing down, $c(T_c) =
0$.] This is true both before and after the transition. That is,
$\bar t$ is defined by the condition that ${\dot\xi}_{ad}(-{\bar
t})\approx c({\bar t})$ or ${\dot\xi}_{ad}({\bar t})\approx -
c({\bar t})$. As a result, ${\bar{t}}$ is of the form
\begin{equation}
{\bar{t}}\sim\tau _{Q}^{1-\gamma }\tau _{0}^{\gamma },
 \label{tbar}
 \end{equation}
where $\tau _{0}\ll \tau _{Q}$ is the cold
relaxation time of the longest wavelength modes, and the critical exponent $%
\gamma $ depends upon the system. It follows that $\tau _{Q}\gg
{\bar{t}}\gg \tau _{0}$.

Domain formation, the frustration of the order parameter fields,
is often visible through topological defects, which mediate
between different equivalent ground states. Since defects are, in
principle, observable, they provide an excellent experimental tool
for confirming this hypothesis when the possibility of producing
them exists. Kibble and Zurek made the further assumption that we can measure ${
\bar{\xi}}$ experimentally by measuring the number of defects,
assuming that the defect separation $\xi _{def}=O({\bar{\xi}})$.
This identification of the initial domain size and defect
separation then gives an estimate of the defect separation at
formation of
\begin{equation}
{\bar{\xi}}\sim \xi _{ad}({\bar{t}})=\xi _{0}\bigg(\frac{\tau _{Q}}{\tau _{0}%
}\bigg)^{\sigma }\gg \xi _{0},  \label{xibar}
\end{equation}
where $\sigma =\gamma \nu $. This is very large on the scale of
cold defects which shrink to size $\xi _{ad}(T_{fin})= O(\xi_0)$,
where $T_{fin}$ is the final temperature. We term $\sigma $ the
Zurek-Kibble (ZK) characteristic index.

In general, we find that the mean-field indices are
\begin{equation}
\gamma = \frac{2}{3},\,\,\,\,\nu = \frac{1}{2};\,\,\,\,\,\,\sigma
= \frac{1}{3}. \label{rel}
\end{equation}
for relativistic systems (e.g. weakly damped quantum fields), and
\begin{equation}
\gamma = \nu = \frac{1}{2};\,\,\,\,\,\,\sigma = \frac{1}{4}
\end{equation}
\label{nonrel} for non-relativistic systems (e.g. superfluids).
There are exceptions to this rule, but we shall not consider them
here.

\section{Stochastic equations}

The question is whether these bounds, independent of the
microscopic equations that govern the phase transition, are
remotely saturated in the physical world. In condensed matter
physics several experiments have been performed to check
(\ref{xibar}). Although the results are mixed the overall
conclusion is positive
\cite{grenoble,helsinki,florence,roberto,technion}.

Experiments cannot be performed for relativistic systems and
models of the early universe, when such transitions were
important, are too ambiguous to be helpful. In practice, an {\it
ab initio} calculation from the microscopic field dynamics
suggests the validity of the scaling laws, but does not permit an
estimate of the efficiency of defect production outside the
framework of mean-field (or large-N) approximations
\cite{stephens,boyanovsky}.

In consequence, a more pragmatic check on the saturation of the
Zurek-Kibble bounds in relativistic systems has been numerical
 \cite{zurek3,lag},
essentially from the empirical damped relativistic Langevin
equation with {\it additive} noise, of the form
\begin{equation}
{\ddot\phi} (x) + \eta{\dot\phi}(x) + (-\nabla^2 +m^2(t)) \phi (x)
+ 2\lambda\phi^3(x)
 = \xi (x),\label{llange}
\end{equation}
where, for simplicity, we consider the theory of a single real
field.
 In Eq.~(\ref{llange}) the time-dependent mass $m(t) = \xi_{ad}^{-1}(t)$,
that triggers the transition, is taken in  the mean-field form
$$m^2(t) = -\mu^2\epsilon
(t),\,\,\,\epsilon (t) = (1-\frac{T(t)}{T_c}),$$ in the vicinity
of  $t=0$, where $\mu^2$ is the cold mass which defines the
symmetry-broken ground states, and $\epsilon (t)\simeq -t/\tau_Q$.
For $t>\tau_Q$ it behaves as $m^2(t)= -\mu^2$. The equilibrium
solution is
$$\phi = \pm v,\,\,\,v^2 = \frac{\mu^2}{2\lambda},$$
the symmetry breaking scale. Numerical simulations \cite{zurek3,lag}
show that the Kibble and Zurek scaling behaviour is recovered in
the limits of small (underdamped) and large (overdamped) $\eta$
respectively, although defects are produced with lower efficiency than
anticipated in (\ref{xibar}).

The overdamped case is the phenomenological non-relativistic
time-dependent Ginzburg-Landau equation
\begin{equation}
\eta{\dot\phi}(x) = -\frac{\delta F}{\delta\phi} + \xi
(x),\label{tdlg}
 \end{equation}
for free energy $F$, often appropriate for condensed matter
systems. Equation (\ref{llange}) arises from {\it linear} coupling
to an environment.

There is, however, a problem with justifying (\ref{llange}), with
its assumption of linear dissipation, for underdamped relativistic
QFT. Although linear couplings have played an important role in
the history of decoherence and Langevin equations, additive noise
alone cannot be justified in QFT, where a pure linear coupling to
the environment corresponds to an inappropriate diagonalisation of
the fields.

More generally, there are two mechanisms for inducing friction
(dissipation) in a relativistic plasma:

a) Changing the dispersion relation of existing particles, as
happens on varying $m^2(t)$, leads to a change in scattering and
decay rates, which leads to a change in the distribution of
particle energies, which leads to friction.

b) The creation of particles from the heat-bath, which leads to
friction.

In each case, we expect dissipative terms of the form
${\dot\phi}\phi^2$ to be important \cite{hosoya,morikawa,lawrie}.
On the other side of the equation there is, equally, a problem
with the noise in (\ref{llange}) with regard to relativistic QFT.
Noise is construed as a consequence of integrating out (tracing
over) environmental degrees of freedom, which are intrinsically
non-linear.
 We assume Boltzmann statistics. In
consequence, the noise, which guarantees that the system is
ultimately driven to its ground states must, by the
fluctuation-dissipation theorem, contain a term of the
multiplicative form $\phi\xi$.

This is confirmed in  linear response theory which, for {\it
small} deviations from equilibrium, leads to Langevin equations of
the form \cite{gleiser,lombmazz,muller,lmr}
\begin{eqnarray}
&&\partial_{\mu}\partial^{\mu} \phi (x) - \mu^2 \phi +
2\lambda\phi^3(x) +\int d^4y
~ K_1(x,y)~ \phi(y)+ \nonumber \\
&&\,\,\,\,+\phi (x) \int d^4y ~ K_2(x,y)~ \phi^2(y)
 =  \xi_2 + \phi (x)\xi_1 (x) + ... .\label{langeqft}
\end{eqnarray}
From the retarded nature of the $K$s a more realistic, albeit
still idealised, equation than (\ref{llange}) for the real
relativistic scalar field $\phi$ is
\begin{eqnarray}
&&{\ddot\phi} (x) + (-\nabla^2 +m^2(t)) \phi (x) +
2\lambda\phi^3(x) + \nonumber \\
&&\,\,\,\,+\eta_2{\dot\phi}(x) +\eta_1{\dot\phi}(x)\phi^2(x)
 = \xi_1(x) + \phi (x)\xi_2 (x),\label{mlange}
\end{eqnarray}
where $\xi_1,\xi_2$ are thermal noise.

We note that, in (\ref{mlange}), the multiplicative noise plays
the role of a stochastic term in the temperature, on rewriting it
as
\begin{eqnarray}
&&{\ddot\phi} (x) + (-\nabla^2 +m^2_{\xi}(t)) \phi (x) +
2\lambda\phi^3(x) + \eta_1{\dot\phi}(x)\nonumber \\
&&\,\,\,\, +\eta_2{\dot\phi}(x)\phi^2(x)
 = \xi_2(x),\label{mlange2}
\end{eqnarray}
where
 $$m^2_{\xi}(t) = -\mu^2\epsilon (1-\frac{T_{\xi}
 (t)}{T_c}),$$
 in which
 \begin{equation}
 T_{\xi}(t) = T(t) - \frac{T_c}{\mu^2}\xi_1(x).
 \end{equation}

 For continuous transitions we are not aware of attempts
 to examine the effects of multiplicative noise for fields (as
distinct from few degree-of-freedom quantum mechanics
\cite{habib,diana,arnold,arnold2,gms,Vdb}) on scaling behaviour,
even at the crudest phenomenological level. [This is not the case
for discontinuous transitions, for which this paper is the
counterpart to \cite{smc}.] We see our work here as a
complementary study to the substantial analysis \cite{zurek3,lag}
that has been undertaken for the simpler and often less believable
equation (\ref{llange}), to which it reduces for small $\eta_2$
and correspondingly weak noise. Finally, although we shall not
pursue this here, we should not see (\ref{mlange}) entirely in the
context of QFT. There has been a considerable effort in condensed
matter to examine multiplicative noise in systems with simple
$\eta{\dot\phi}$ dissipation, but with $\xi\phi$ multiplicative
noise ($\eta_2 = \xi_2 = 0$) \cite{sakaguchi,becker}. This leads
to non-Boltzmann (Tsallis/Renyi) distributions
\cite{tsallis,renyi}. Multiplicative noise can both induce phase
transitions and restore symmetry .

\section{Numerical Simulations}

It is sufficient for our purposes to restrict ourselves to a real
field on the line (i.e. 1+1 dimensions), for which $\nabla^2 =
\partial^2/\partial x^2$. It was for such a system that the original
equation (\ref{llange}) was solved \cite{lag}. In that case
extension to higher dimensions gave few new insights as to the
Kibble mechanism, and we expect the same here.

Defects in this case are kinks,
\begin{equation}
\phi_{\mathrm{kink}} (x)= \pm v \tanh\left(\frac{\mu}{2\sqrt{2}}x\right)
\end{equation}
of thickness $\xi_0=\mu^{-1}$ and energy $E = O(\mu^3/\lambda)$.
Although, rigorously, there are no transitions for such
short-range interactions in 1+1 dimensions, there is an effective
transition for medium times. Typically, some time after the end of
the quench, the field settles into a set of alternating
positive/negative vacuum  regions. These are separated by well
defined kinks/anti-kinks that evolve slowly, possibly annihilating
each other for very long times. In this regime defects coincide
with the zeroes of the field, making it straightforward to
identify them in a numerical setting. Clearly this procedure  is
ambiguous for very early times, since zeroes occur at all scales,
and only some of these will evolve into the cores of kinks.
 Here we will restrict ourselves to looking
at kinks {\it after} the quench has terminated, thus avoiding any
counting ambiguities.

\subsection{Numerical Setup}

To further clarify the possible effects of multiplicative noise
terms in the mechanism of defect formation, we performed a
numerical study of the model described in (\ref{mlange}). The
approach followed is close to that in Ref.\cite{lag} for the case
of a quenched one-dimensional system with purely additive white
noise. We evolve the following 1+1 Langevin equation:
\begin{eqnarray}
&&\ddot{\phi}-\nabla^2 \phi
+[ \alpha_1^2 \phi^2 + \alpha_2^2 ]
\,\dot{\phi}-m^2(t) \phi - 2 \lambda \phi^3 \nonumber \\
&&\,\,\,\,=\alpha_1 \phi \xi_1 + \alpha_2 \xi_2
\label{eom_num}
\end{eqnarray}
where $m^2(t)=-\mu^2\epsilon(t)$, $\mu^2=1.0$ and $\lambda=1.0$.
$\xi_1$ and $\xi_2$ are uncorrelated gaussian noise terms obeying
\begin{eqnarray}
 && \langle\xi_a(x',t') \xi_b(x,t)\rangle=2 T \delta_{a b}
     \delta(x'-x) \delta(t'-t), \nonumber \\
 && \langle \xi_a(x,t) \rangle=0 .
\end{eqnarray}
The bath temperature $T$ is set to a low value, typically
$T=0.01$.  The relative strength of the multiplicative and additive noise is
measured by $\alpha_1$ and $\alpha_2$, with corresponding dissipation terms
obeying the fluctuation-dissipation relation. The values of $\alpha_1$ and
$\alpha_2$ vary between different sets of runs, allowing us to compare
their effects on the final defect density.

 When dealing with stochastic equations with multiplicative noise
 one usually has to take into account that the continuous equation
 may not have a unique interpretation. This is the well known
 It\^{o}-Stratanovich ambiguity \cite{gardiner} which is usually
 resolved by singling out a specific discretisation of the equations
 of motion. It turns out that this problem has no relevance for the
 type of system we are considering. This could be seen by obtaining
 explicit Fokker-Plank equations for different time-discretisations
 of the model above. Since the multiplicative term depends only on
 the field (an not on its time derivative), the Fokker-Plank equation
 turns out to be the same for all alternative interpretations
 \cite{habib,arnold2}.

 The equations of motion were discretised using a second-order leap-frog
 algorithm. We set  $\delta x=0.125$ and $\delta t=0.1$
in a periodic simulation box with $8000$ to $16000$ points. Note that
the core of the defect (with finite temperature size $1/m$)
is resolved by 8 lattice points which should be enough for our
purposes.

 Walls are identified by looking at zero crossings of the
scalar field. As discussed above, this should be accurate for long
times and low values of $T$ - precisely the regime where we
measure the final values for the defect density. For each
individual quench, the final number of defects is determined by
counting kinks at a final time, defined as a fixed multiple of the
quench time-scale. There are more complex ways of defining a final
defect density, namely by fitting the time-dependence of the kink
number to exponential or power-law decay expressions. The several
approaches were compared in \cite{lag}. The conclusion was that
the straightforward kink counting performs less well in the very
fast/slow quench-time limits, leading to slightly higher estimates
of $\sigma$.
 With this caveat in mind we will keep to the simpler
 approach since, as we will see, it is accurate enough to illustrate well the
 effects of multiplicative noise terms.

 For every fixed choice of $\alpha_1$ and $\alpha_2$  we perform a series of quenches,
with the quench times varying as $\tau_Q=2^n$,  $n=1,2,..,9$. Each
quench is repeated several times (typically 10) with different random number
realizations, and the final defect number is averaged over this ensemble.
The scaling exponent $\sigma$ can then be obtained by fitting the final defect
number dependence on $\tau_Q$ to a power-law of the form $A \tau_Q^{-\sigma}$.

\subsection{Simulation Results for Kink Densities}

\begin{figure}
\scalebox{0.50}{\includegraphics{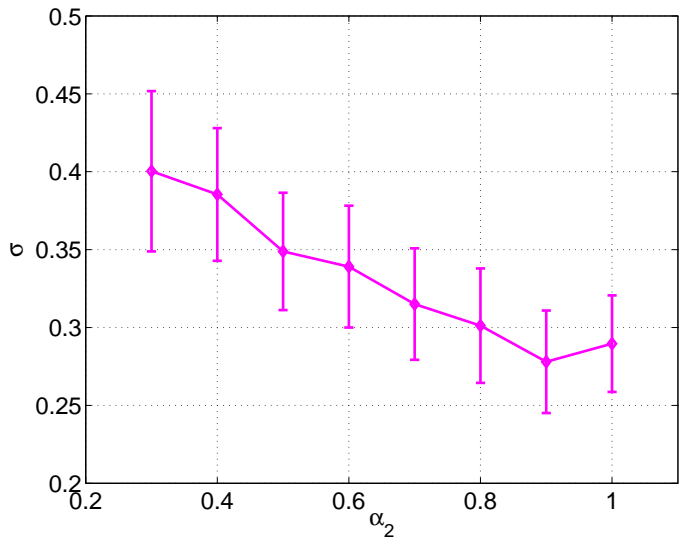}} \caption{Final defect
density scaling power as a function of the additive noise
strength. The multiplicative noise term is set to zero. Error bars
represent the standard deviation of the result over 10 independent
series of quench realizations.} \label{additive}
\end{figure}

 We start with the simple purely additive noise case, corresponding to
setting $\alpha_1=0$ in Eq.~(\ref{eom_num}). In Fig.\ref{additive}
we can see the dependence of the scaling power $\sigma$, in terms
of noise strength $\alpha_2$. Our results are very similar to
those found in \cite{lag}, with $\sigma$ decreasing as the value
of the dissipation, $\eta=\alpha_1^2$ increases. This takes us
from the relativistic regime where we expect $\sigma\simeq1/3$ to
the overdamped case with $\sigma\simeq1/4$. As observed by
previous authors, for very small values of the dissipation, the
scaling fails to follow the power-law rule, a consequence of
saturation. This explains the high, un-physical values of $\sigma$
for $\alpha_2<0.5$ (corresponding to a dissipation of
$\eta=0.25$). The quality of the power-law fit becomes poor in
this parameter region, a further sign of deviations from the
simple scaling behaviour.

\begin{figure}
\scalebox{0.5}{\includegraphics{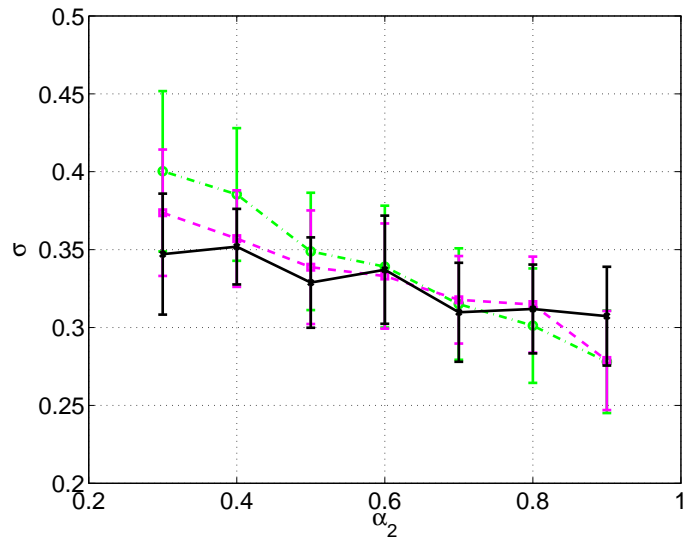}}
\caption{Final defect density power versus additive noise strength. The
 multiplicative noise term is $\alpha_1=0, 0.5$ and $0.9$  for the green
 (dahsdot), pink (dashed) and black (solid) curves respectively.
 Error bars as before.}
\label{additive_multiplicative}
\end{figure}

Next we look at how the introduction of multiplicative noise
influences the results. In Fig.\ref{additive_multiplicative} we
have the scaling power for $\alpha_2$ in the same region as
Fig.\ref{additive}, for three different values of multiplicative
noise strength, $\alpha_1=0.0, 0.5$ and $0.9$ respectively.
Clearly the results change very little - within error bars, the
three curves basically overlap each other.

\begin{figure}
\scalebox{0.50}{\includegraphics{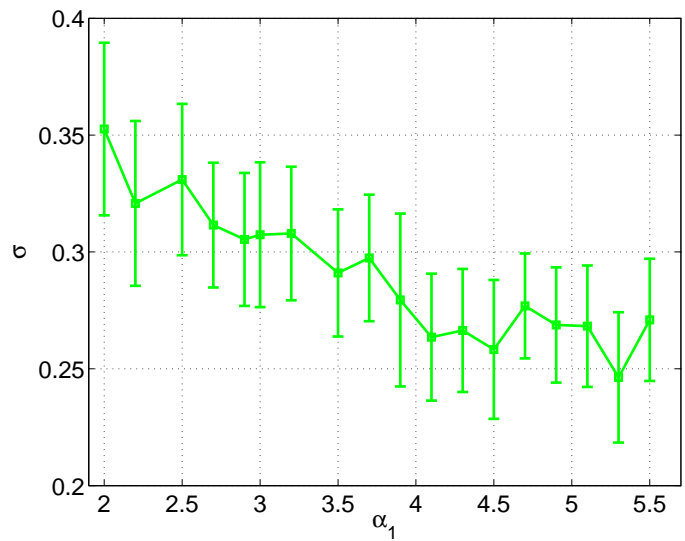}} \caption{Defect
density power as a function of multiplicative noise strength
$\alpha_1$. The additive noise is set to zero. Error bars as
before.} \label{multiplicative}
\end{figure}

Though this result
may seem disappointing at first, we should be aware that the significant range
of variation of $\alpha_1$ may differ considerably from that for $\alpha_2$.
This is illustrated in Fig.\ref{multiplicative} where we have $\sigma$
for higher values of $\alpha_1$, with the contribution of the additive
term set to zero. The pattern is the same as observed before for the
case of additive noise. As $\alpha_1$ increases, the scaling exponent changes
from relativistic values to those typical of an overdamped
system. The transition between the two types of behaviour takes place
for values of $\alpha_1$ considerably higher than $\alpha_2$.
 Defining the noise strength transition value as
the one above which $\sigma<0.3$, we have $\alpha_2\simeq 0.8$ for
purely additive noise and $\alpha_1\simeq 3.5$ in the multiplicative case.

\begin{figure}
\scalebox{0.50}{\includegraphics{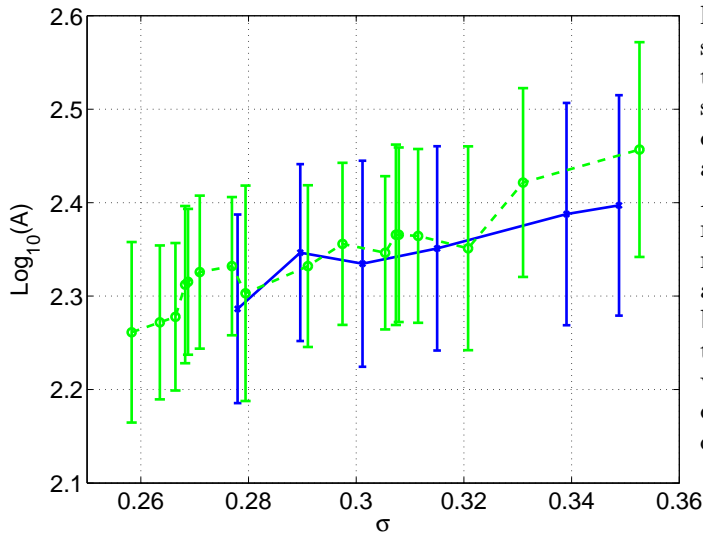}} \caption{Amplitude of
the power-law in terms of the power $\sigma$ for pure
multiplicative (green dashed plot) and purely additive (blue solid) noise.
Error bars as before.} \label{amplitude}
\end{figure}

 These results can be understood if we make the simple assumption
that the order of magnitude of the effective dissipation in the
multiplicative noise case is given by $\eta\simeq\alpha_1^2
\langle \phi^2 \rangle$ \cite{habib}, the mean of the term
multiplying $\dot{\phi}$ in Eq.~\ref{mlange}. $\langle \phi^2
\rangle$ is the average of the square of the field during the
stage of the quench determining the scaling, that is, slightly
before the critical temperature is reached. If the typical value
of $\langle \phi^2 \rangle$ is small, the effective dissipation
for multiplicative noise should be reduced. As a consequence,
relatively high values of $\alpha_1$ should be required to take
the system from the relativistic to the dissipative regime.

  This argument can be made more quantitative by noting that in the
purely additive case the under-damped to over-damped transition
takes place for $\eta\simeq0.8^2\simeq0.6$. A similar effective dissipation
with pure multiplicative noise would be reached for $\alpha_1=3.5$ if
 $\langle \phi^2 \rangle\simeq0.6/3.5^2 \simeq0.05$.
 We measured the value of the mean squared field explicitly in the
simulations, and observed that at $t=0$ one has typically $\langle
\phi^2 \rangle\simeq0.01 - 0.02$, with the higher value
corresponding to the slower quenches. This is indeed of the same
order of magnitude as required by the above reasoning. We note
that, if instead, we were to replace $\alpha_1^2 \langle \phi^2
\rangle$ by $\alpha_1^2v^2$ we would obtain an effective dissipation
of $\eta=6.1$, one order of magnitude larger than the critical value.
This suggests that even though kinks can only be identified with
rigour once $$\langle\phi^2\rangle\sim v^2,$$ the period of the
evolution responsible for setting the relevant defect separation
scale takes place considerably earlier, in accord with
(\ref{tbar}).

 The above result can be extended to systems with both kinds of noise,
 with the generic effective dissipation being given by $\eta=\alpha_2^2 +
\alpha_1^2 \langle \phi^2 \rangle$. For $\alpha_1<1.$ the
correction to the multiplicative component should be less than
$0.05$. This explains why the inclusion of multiplicative noise of
this magnitude changes very little the additive noise result, as
illustrated in Fig.\ref{additive_multiplicative}.

Finally we checked whether the inclusion of multiplicative noise
leads to any appreciable change in the behaviour of the amplitude
of the power-law. In Fig.\ref{amplitude} we show the value of
$\log(A)$ where $A$ is the amplitude of the fit for the final
defect density $A \tau_Q^{-\sigma}$. The results are shown as a
function of the power
 $\sigma$, corresponding to different ranges of $\alpha_1$ and $\alpha_2$
in the purely multiplicative and additive noise cases
respectively. As can be observed, $A$ does not differ
significantly between the two types of noise, for similar values
of $\sigma$. Physically this implies that not only the scaling
power is similar in both cases, but also the overall amount of
defects produced is not affected by the type of noise involved in
the transition. Overall the conclusion seems to be that once we
adjust for the value of the effective dissipation, the
distribution of the defect number in the final configuration is
independent of the properties of the noise terms driving the
system.

\section{Conclusions}

Since most effective equations of motion derived from field theory
involve both additive and multiplicative noise, it is natural to
wonder whether the presence of multiplicative noise changes the
dynamics of non-equilibrium phase transitions. Here we looked in
detail at the formation of defects in a $1+1$ dimensional system
undergoing a quench with both types of thermal noise present.
Surprisingly, we found that the properties of the defect
population after the quench can be well described in terms of the
Kibble-Zurek scenario, if we take into account the effects of
multiplicative noise in the dissipation. In particular,
multiplicative noise terms increase the dissipation by an amount
of the order of the noise amplitude, times the value of the mean
field square at the transition time. As in the purely additive
case, we observe that for low values of the effective dissipation
the system behaves in a relativistic fashion, with the final
number of defects scaling as a power of $1/3$ of the quench time.
For higher values the system enters an over-damped regime
characterized by a lower scaling power, nearer to $1/4$. Although
our multiplicative noise was the most simple (tracing over short
wavelength modes will also induce $\phi^2\xi_3$ noise) this
suggests that the Kibble-Zurek scaling laws are robust.

It is interesting to note that these results still leave open the
question of whether the defect density (or equivalently, the freeze-out
correlation length) is set before or after the transition takes place,
i.e. for $\bar{t}<0$ or $\bar{t}>0$. Strictly, the value for the average
field square $\langle\phi^2\rangle$ used in defining the effective
dissipation, should
be evaluated at $\bar{t}$. Unfortunately the value of this quantity varies
slowly in the vicinity of the critical point and as a consequence it is
not possible to determine whether the correct value is fixed above or
below $T_c$. We can nevertheless, use our model as a setting for answering
this question, by performing quenches where the value of the effective
dissipation is forced to change at $t=0$. By looking at the defect
scaling in cases where the shift in the dissipation takes the system
from an under-damped to an over-damped regime at $t=0$, the significant
stage of the evolution should become apparent.
We will clarify these points in detail in a future publication.

\section{Acknowledgements}

 N. D. Antunes was funded by PPARC. P. Gandra was supported by FCT,
 grant number PRAXIS XXI BD/18432/98. RR would like to acknowledge support
 from the ESF COSLAB programme.

\end{document}